\documentclass{emulateapj}
\usepackage{float}
\usepackage{amsmath}
\usepackage{epsfig,floatflt}
\usepackage{subfigure}



\usepackage{amsmath}
\usepackage{epsfig}
\newcommand{\beq}{\begin{equation}}
\newcommand{\eeq}{\end{equation}}
\newcommand{\be}{\begin{eqnarray}}
\newcommand{\ee}{\end{eqnarray}}

\def\cmb{{\rm c}}
\def\ilc{{\rm ilc}}
\def\fg{{\rm F}}
\def\wmap{{\sl WMAP }}

\def\healpix{H{\sc ealpix }}
\def\glesp{G{\sc lesp }}
\def\l{{\ell}}
\def\xcf{{X_{\cmb \fg}}}

\def\alm{a_{\l m}}

\def\cl{C_{\l}}

\def\lm{{\l m}}

\def\f{{\rm fr}}
\def\dtsl{{\delta T_\l^2}}

\def\etal{{et al.}}

\slugcomment{Submitted to the {\it Astrophysical Journal}}

\begin{document}
\newcommand{\asiaa}{{Institute of Astronomy and Astrophysics, Academia Sinica, P.O. Box 23-141, Taipei 10617, Taiwan, R.O.C.}}
\newcommand{\cardiff}{{School of Physics \& Astronomy, Cardiff University, 5 The Parade, Cardiff, CF24 3AA, Wales, United Kingdom}}
\newcommand{\nbi}{{Niels Bohr Institute, 17 Blegdamsvej, Copenhagen, DK-2100, Denmark}}
\newcommand{\sfu}{{Space Research Department, Southern Federal University, Zorge, 5, 344091, Rostov on Don, Russia}}

\altaffiltext{1}{\asiaa}
\altaffiltext{2}{\nbi}
\altaffiltext{3}{\sfu}
\altaffiltext{4}{\cardiff}

\title{Cosmic Covariance and the low Quadrupole Anisotropy of the Wilkinson Microwave Anisotropy Probe (WMAP) data}
\author{
Lung-Yih Chiang\altaffilmark{1},
Pavel D. Naselsky\altaffilmark{2,3},
Peter Coles\altaffilmark{4}
}

\email{lychiang@asiaa.sinica.edu.tw, naselsky@nbi.dk, peter.coles@astro.cf.ac.uk}

\keywords{cosmology: cosmic microwave background --- cosmology:
observations --- methods: data analysis}

\begin{abstract}

The quadrupole power of cosmic microwave background (CMB)
temperature anisotropies seen in the \wmap data is puzzlingly low.
In this paper we demonstrate that Minimum Variance Optimization
(MVO), a technique used by many authors (including the \wmap science
team) to separate the CMB from contaminating foregrounds, has the
effect of forcing the extracted CMB map to have zero statistical
correlation with the foreground emission. Over an ensemble of
universes the true CMB and foreground are indeed expected to be
uncorrelated, but any particular sky pattern (such as the one we
happen to observe) will generate non-zero measured correlations
simply by chance. We call this effect ``cosmic covariance'' and it
is a possible source of bias in the CMB maps cleaned using the MVO
technique. We show that the presence of cosmic covariance is
expected to artificially suppress the variance of the Internal Linear
Combination (ILC) map obtained via MVO. It also propagates into the
multipole expansion of the ILC map, generating a quadrupole
deficit with more than 90\% confidence. Since we do not know the CMB
and the foregrounds {\it a priori}, there is therefore an unknown
contribution to the uncertainty in the measured quadrupole power,
over and above the usual cosmic variance contribution. Using the MVO
on a series of Monte Carlo simulations that assume Gaussian CMB
fluctuations, we estimate that the real quadrupole power of the CMB
lies in the range $[305.16,400.40] (\mu {\rm K}^2)$ (at the
$1-\sigma$ level).

\end{abstract}

\section{Introduction}

Detailed measurements of the pattern of temperature anisotropies in
the cosmic microwave background (CMB) have provided cosmologists
with unprecedented opportunities to probe the physics of the early
Universe. In particular, the Wilkinson Microwave Anisotropy Probe
(WMAP) data \citep{wmapresults,wmapfg,wmapcl,wmap3ytem} have played
a pivotal role in the establishment of the standard ``concordance''
cosmological model and have heavily constrained alternatives to the
inflationary paradigm for the origin of large-scale structure.

According to the concordance cosmology, the variations in
temperature of the CMB across the celestial sphere should possess a
very broad spectrum usually expressed in terms of the amplitude of
spherical harmonic modes labelled by the usual angular harmonic
frequency $\l$. The power of the pattern of anisotropies power is
expected to be strongest for the quadrupole ($\l=2$) which
correspond to variations on an angular scale of  $90^\circ$ on the
sky. In principle, therefore, the quadrupole should be the easiest
harmonic mode to detect. In practice, however, the measured
quadrupole anisotropy of the CMB from \wmap seems to sit rather
uncomfortably with the standard cosmology: its amplitude seems too
low \citep{efssig,dtzh}. Of course, the power that resides in each multipole of the CMB
pattern (including the quadrupole) is not so much measured as
extracted. There are various diffuse foreground emissions to be
separated from the ``true'' primordial CMB, such as free-free
emission arising from electron-ion scattering, synchrotron emission
from cosmic ray electrons accelerating in the galactic magnetic
fields, and dust emission. However, although a number of authors
have  employed different foreground subtraction schemes, all find
the derived quadrupole power to be lower than the Concordance
Cosmological Model with a significance level of 5\% or better
\citep{efssig,lilc,dt,park3y,saha}. This has even prompted suggestions of
physics beyond the standard model \citep{efsspacurv,dodecahedral}.

In this paper we suggest that the quadruple power deficit in \wmap
may be an artifact of the Minimum Variance Optimization method
used by many authors to clean CMB maps from foreground
contamination. In the next Section, we explain the basics of the MVO
method. In Section 3 we show how this method is prone to a bias
introduced by the accidental alignment of structures in the CMB and
foreground templates leading to an effect we call ``cosmic
covariance''. In Section 4 we quantify the likely effect of this
bias using Monte Carlo simulations, and we estimate the quadrupole power taking into account the effect of cosmic covariance. We present our
conclusions in Section 6.

\section{Minimum Variance Optimization}
Although various schemes have been proposed for subtracting
foregrounds from CMB data, one concept that tends to be in common
for multi-frequency cleaning methods is Minimum Variance
Optimization (MVO). The theoretical basis for MVO is the primordial
CMB temperature anisotropies arise from black-body radiation and
therefore constitute a frequency--independent signal that persists
across different observed frequency bands \citep{te96}. The observed sky
temperature in a given band $i$ is therefore just
\begin{equation}
T_i = T_\cmb + f_i,
\end{equation}
where $T_\cmb$ is the CMB signal and $f_i$ the foreground emission
band $i$, which will be a composite of dust, free--free and
synchrotron sources. For the purposes of this paper we assume that
instrument noise is negligible. The frequency-independent signal can
be flushed out by an Internal Linear Combination (ILC) from the maps
of $M$ frequency bands with weighting coefficients $w_i$, where
$\sum_{i=1}^M w_i=1$ \citep{wmapfg}:
\begin{equation}
T_\ilc(p)=\sum_{i=1}^M w_i T_i(p) = T_\cmb(p) + \sum_{i=1}^M w_i
f_i(p)\equiv T_\cmb(p)+T_\f(p), \label{minivar}
\end{equation}
where $T_\ilc(p)$ denotes the ILC map value at a pixel $p$ and $T_\f$ the foreground residual. In
general, 
\begin{equation} 
{\rm Var} (T_\ilc) = {\rm Var}(T_\cmb) + {\rm
Var}(T_\f) + 2 {\rm Cov} (T_\cmb, T_\f), \label{cov} 
\end{equation}
where the variances and covariance are measured over an ensemble of
possible skies. The MVO technique is based on the {\it a priori}
assumption that the frequency-independent signal (i.e. the true CMB)
should be {\it statistically} independent of any foreground
contribution, so that the covariance term vanishes and 
 \begin{equation}
{\rm Var} (T_\ilc) = {\rm Var}(T_\cmb) + {\rm Var}(T_\f),
\end{equation}
or in other words,
\begin{equation}
\langle \sigma_\ilc^2\rangle= \langle \sigma_\cmb^2 \rangle + {\rm
Var} \left[ \sum_{i=1}^M w_i f_i(p) \right], \label{si}
\end{equation}
where the angle brackets denote ensemble averages as above.
Minimizing the variance of the ILC map is then equivalent to
minimizing the foreground contamination of a map of thermal noise.

The \wmap 3-year ILC map and, more importantly, the power spectrum for $\l\le30$ are\footnote{For power spectrum for $\l > 30$, \wmap science team applies a $\chi^2$ foreground template fitting method.} thus produced by employing the MVO on 5 frequency maps K (23 GHz), Ka (33 GHz), Q (41 GHz), V (61 GHz) and W (94 GHz) bands in
the pixel domain for 12 separate regions, where Region 0 marks the
largest region with $|b| \geq 15^\circ$ (about 89\% of the whole
sky), and Regions $1-11$ are those around the Galactic plane. For
each region the 5 frequency band maps are linearly combined and a
set of weighting coefficients $w_i^{(R)}$ for each region are
obtained in such a way that the resultant variance is forced to be
minimum:
\begin{equation}
\frac{\partial \sigma_\ilc^2}{\partial w_i} =
\frac{\partial}{\partial w_i} {\rm Var} \left[ \sum_{i=1}^M w_i
f_i(p) \right]=0.
\end{equation}
Note that Equation (\ref{si}) involves the requirement of
statistical independence over an ensemble of universes. As we have
only one universe (i.e. one CMB sky), the calculation of variances
and covariances can only be done over the set of pixels
corresponding to a single sky, and not over an entire probably
distribution. The sky covariance between $T_\cmb$ and $T_\f$ need not
be exactly zero, since there is always a chance of some chance
coincidence of features such as hotspots between foreground and true
CMB. If this happens then
\begin{equation}
\frac{\partial \sigma_\ilc^2}{\partial w_i} \ne
\frac{\partial}{\partial w_i} {\rm Var} \left[ \sum_{i=1}^M w_i
f_i(p) \right].
\end{equation}
As we shall now show, this leads to a bias in the recovery of the
CMB map.

\section{Cosmic Covariance}

Let us assume (for illustrative purposes only) that the components
of the foreground emission at different frequencies have the same
spatial distribution (i.e uniform foreground spectra). In reality
the foregrounds at different frequency bands have spatial variations
of the spectral indices, which can amplify the resulting bias. In
the uniform case we have $f_i(p) = S_i F(p)$, where $S_i \equiv
S(\nu_i)$ is the composite frequency spectrum of the foreground
emission and $F$ is the common spatial distribution over the set of
pixels. Thus frequency band maps $T_i(p)= T_\cmb(p) + S_i F(p)$ and the ILC map
\begin{equation}
T_\ilc(p) = T_\cmb(p) + \Gamma F(p),
\end{equation}
where $\Gamma \equiv \sum_i^M w_i S_i$. The variance of the ILC map can be written as follows
\citep{wmap3ytem}
\begin{equation}
\sigma_\ilc^2=\langle T_\ilc^2(p) \rangle - \langle T_\ilc(p)
\rangle^2 = \sigma_\cmb^2 +2 \Gamma \sigma_{\cmb, \fg} +\Gamma^2
\sigma_\fg^2, \label{combin}
\end{equation}
where 
\begin{equation}
\sigma_{\cmb, \fg}\equiv \langle T_\cmb F \rangle - \langle
T_\cmb \rangle\langle F\rangle 
\end{equation}
is the ``cosmic covariance''
between the CMB and the foreground spatial distribution $F$. In this
and the subsequent equations, the angle brackets denote averages
over the pixels of a single sky rather than ensemble averages; c.f.
Equation (\ref{si}).

The term ``cosmic covariance'' is coined similarly to ``cosmic
variance'' \citep{cosmicvariance} because both effects emanate from the fact that we must
do our statistical analysis using a single version of the celestial
sphere. Cosmic variance arises from the fact that a single sky does
not offer sufficient spherical harmonic modes to make an accurate
estimate of the ensemble-averaged power at low multipoles. In the case of cosmic covariance, the single sky prevents us from being sure that our
calculation of the correlation between foreground and CMB is
exact, particularly for the {\it quadratic minimization}. The value $\sigma_{\cmb, \fg}$ will be non-zero merely
because the CMB and the foregrounds happen to line up to a certain
degree simply by chance. If both CMB and foreground contain
large-scale structures then the effect of cosmic covariance will be
larger as the sky will contain fewer independent regions.

The cosmic covariance $\sigma_{\cmb, \fg}$ can also be expressed via
the correlation coefficient $\xcf$: $ \sigma_{\cmb, \fg} \equiv \xcf
\sigma_\cmb \sigma_\fg$, where $\xcf$, the ``cosmic correlation
coefficient'' characterizes the level of resemblance in morphology
between the CMB and the foreground spatial distribution, and $-1 \le
\xcf \le 1$.

Employing the MVO criterion to the variance of the ILC map, we get
\begin{equation}
\frac{\partial \sigma_\ilc^2}{\partial w_i} = 2 \frac{\partial \Gamma}{\partial w_i}
\xcf \sigma_\cmb \sigma_\fg +2 \Gamma \frac{\partial \Gamma}{\partial w_i} \sigma_\fg^2 =0,
\end{equation}
which gives $\Gamma = -\xcf \sigma_\cmb/\sigma_\fg$ and the ILC map is then
\begin{equation}
T_\ilc (p)= T_\cmb (p) - \frac{\xcf \sigma_\cmb }{\sigma_\fg} F(p).
\label{ilcmap}
\end{equation}
One can see if cosmic covariance is zero (so is $\xcf$), then $T_\ilc
(p)= T_\cmb (p)$.

The last term $-\xcf  F(p)\sigma_\cmb /\sigma_\fg $ is usually
discussed in terms of an  anti-correlation bias \citep{wmap3ytem},
but it is in fact due to the MVO serving to eliminate the covariance
between the ILC and the foreground spatial distribution. To see
this, note that the covariance between the ILC and the foreground
is forced to vanish:
\begin{equation}
\sigma_{\ilc, \fg}\equiv \langle T_\ilc F \rangle - \langle T_\ilc
\rangle\langle F\rangle =\sigma_{\cmb, \fg}- \frac{\xcf
\sigma_\cmb}{\sigma_\fg} \sigma_\fg^2 = 0.
\end{equation}
The ILC map is thus guaranteed to have a morphology that has no
resemblance to that of the foregrounds, regardless of whether or not
the CMB has any such resemblance. That is to say, even if there
exists any cosmic covariance between the true CMB signal and the
foregrounds, it is duly subtracted by the MVO, and the resulting ILC
map must display zero sky correlation with the foregrounds.

Forcible subtraction of the unknown cosmic covariance also affects
the overall variance.  Putting $\Gamma = -\xcf
\sigma_\cmb/\sigma_\fg$ into Eq.(\ref{combin}), we have
\begin{equation}
\sigma_\ilc^2=\sigma_\cmb^2 (1-\xcf^2).
\end{equation}
Subtraction of the cosmic covariance manifests itself as a deficit
in the total variance of a factor $(1-\xcf^2)$. This can partially account for the low variance in the CMB map \citep{lowvar}. As we do not know
$T_c$ and $F$ {\it a priori}, we are left with no information about
the level of cosmic covariance after it is subtracted.

One of the dangers of the MVO approach is that it encourages
circular reasoning. The {\it a priori} assumption is that the CMB
and the foregrounds are statistically independent. The MVO produces
an ILC map that has zero sky correlation with the foregrounds. This
seems to confirm the correctness of this assumption.  However, the
MVO must produce this result whatever the morphology of the input
templates.

\begin{table}
\begin{center}
\begin{tabular}{|c|c|c|c|c|c|} \hline
      &  K      &  Ka   &  Q       &  V     &  W      \\  \hline
K     &         & 0.9986 & 0.9968  & 0.9964 & 0.9778  \\  \hline
Ka    & 0.9987  &        & 0.9994  & 0.9989 & 0.9749  \\  \hline
Q     & 0.9966  & 0.9994 &         & 0.9991 & 0.9720  \\  \hline
V     & 0.9957  & 0.9985 & 0.9993  &        & 0.9774  \\  \hline
W     & 0.9835  & 0.9790 & 0.9756  & 0.9790 &         \\  \hline
\end{tabular}
\end{center}
\caption{Correlation coefficients $X$ between the full-sky \wmap
foreground maps (lower left triangle) and between those in the
\wmap-defined Region 0, which accounts for 89\% of the full sky
(upper right triangle). One can see that K, Ka, Q and V channels
(except W) have correlation very close to unity. All other Regions
of the \wmap channels except W show the same trend.}
\label{xcorrtable}
\end{table}

Although the above analysis is based on the existence of composite
foregrounds with uniform frequency spectra, we can see that it is
not far from reality. In Table 1, we show the pixel-by-pixel
correlation coefficients between the \wmap composite foregrounds in
{\it WMAP}-defined Region 0, which contains $\sim 89$\% of the whole
sky (upper right triangle). One can see that all channels have
excellent agreement in terms of morphology (with the exception of W
channel due to synchrotron emission). The lower left triangle part shows
correlations between full-sky foreground maps, which also display the
same trend.

\section{Monte Carlo Simulations}

To demonstrate the power deficit and reduction of cosmic covariance
in the ILC map due to the MVO, we perform simulations of 2000 sets
of frequency band maps corresponding to the \wmap K, Ka, Q, V and W
channels. These comprise the sum of a simulated CMB signal (assuming
Gaussian fluctuations), mock foreground maps and instrument noise
maps. The CMB signal is simulated with the \wmap best-fit
$\Lambda$CDM model. The mock foreground maps at \wmap 5 frequency
bands are produced as follows: we take the \wmap composite
foreground at K band, which is multiplied by a factor
$\rho_i$ such that $\rho_i \sigma_{\rm K}= \sigma_i, i= $ Ka, Q, V
and W where $\sigma_i$ is the standard deviation of the \wmap
composite foreground maps (in thermodynamic temperature units). As
such, these mock foreground maps have exactly the same morphology,
but each has the same variance as the \wmap foreground map of the
corresponding frequency band. The Gaussian instrument noise map is
with $\cl= 2 \times 10^{-8}$ (${\rm mK}^2$). To follow the \wmap procedure, both the simulated CMB and noise maps are smoothed to $1^\circ$ FWHM before summation with the mock foregrounds (which are already smoothed). The MVO is then employed on the internal
linear combination of the band maps to retrieve the CMB signal.

\begin{figure}
\plotone{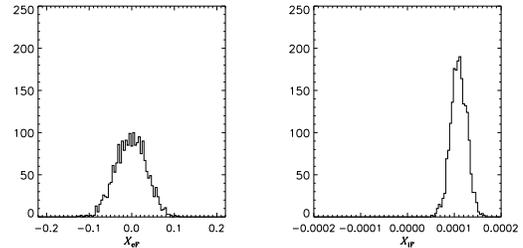}
\caption{Decreasing of correlation with the foreground due to the MVO. We show the histogram of $X_{\rm cF}$ and $X_{\rm iF}$, the correlation coefficient between the foreground and the simulated CMB (left) and the retrieved CMB (right) via the MVO.}
\label{xcorr}
\end{figure}

In Figure \ref{xcorr} we plot the histogram of $\xcf$ (left), the
correlation coefficient between the simulated CMB and the common foreground
map, and that of $X_{\rm iF}$ (right), the correlation coefficient
between the retrieved CMB (ILC) and the common foreground map. One can see
the sky correlation is reduced below a level $2 \times 10^{-4}$
after the application of the MVO.

The deficit in the variance $\xcf^2 \sigma_c^2$ is registered in the
angular power spectrum $\cl$ via $\sigma^2_\ilc= (4\pi)^{-1} \sum_\l
(2\l+1) \cl^\ilc$. Note that $\cl$ is always positive, so the
deficit in variance manifests itself as a few dips compared to the
desired CMB angular power spectrum $\cl^\cmb$. Since the MVO acts on
the overall variance, it only ensures the variance is minimum; the
result for individual spherical harmonic modes is less obvious.
Decomposing Eq.(\ref{ilcmap}) into spherical harmonic coefficients
$\alm^\ilc= \alm^\cmb- X\sigma_\cmb \alm^\fg/\sigma_\fg$, we can
examine its angular power spectrum
\begin{eqnarray}
\cl^\ilc&=&\frac{1}{2\l+1}\sum_{m=-\l}^{\l} |\alm^\ilc|^2 = \cl^\cmb + \frac{X^2\sigma_c^2}{\sigma_\fg^2} \cl^\fg \nonumber \\
&-&\frac{2X\sigma_\cmb}{\sigma_\fg (2\l+1)} \sum_{m=-\l}^{\l} |\alm^\cmb| |\alm^\fg|\cos(\phi^\cmb_\lm-\phi^\fg_\lm).
\end{eqnarray}
One can see that it is possible for some multipoles to acquire
excess power (bumps) over  $\cl^\cmb$ if, for instance, $ \pi/2
<\phi^\cmb_\lm-\phi^\fg_\lm < 3\pi/2$. So in order to preserve the
overall deficit,  it shall drag the dips
at other multipoles even further down.

\begin{figure}
\plotone{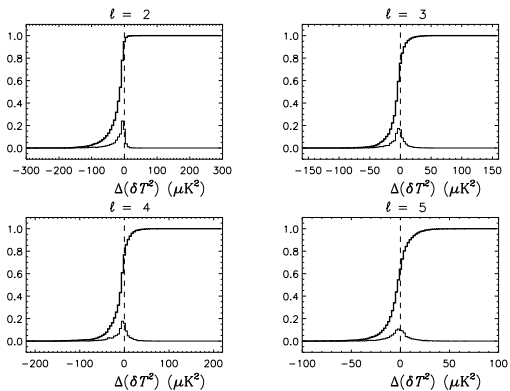}
\caption{Normalized histograms (thin) and cumulative histograms (thick) of the power difference $\Delta (\dtsl) \equiv \dtsl_{\rm i}-\dtsl_{\rm s}$ from 2000 realizations for $\l=2$, 3, 4 and 5, where $\dtsl_{\rm s}$ denotes the simulated CMB power at $\l$ mode and $\dtsl_{\rm i}$ the retrieved CMB (ILC) power. The MVO is employed on frequency band maps with the foregrounds having the same morphology (see text for the simulations). One can see that the CMB retrieved via MVO has a skew distribution for $\l=2$, indicating the power of the retrieved CMB is smaller than the simulated ones.}
\label{deficit}
\end{figure}

In Figure \ref{deficit} we plot the normalized histogram and
cumulative histogram of the power difference  $\Delta (\dtsl) \equiv
\dtsl_{\rm i}-\dtsl_{\rm s}$, where $\dtsl =  \l (\l+1) \cl/(2\pi)$ is the
temperature anisotropy power at multipole $\l$, the subscripts ${\rm s}$
and ${\rm i}$ denote the simulated CMB and the retrieved CMB from MVO,
respectively. It is easy to see that, particularly for multipole number $\l=2$, the simulation ensemble forms a skewed distribution with a tail on the negative side,
indicating the power of the retrieved CMB is more likely to be
smaller than that of the input (simulated) CMB. For other multipoles the skew is less obvious. This shows, therefore, that the deficit in variance due to the MVO is most likely registered in the quadrupole. 

One should note that the power deficit claimed by the \wmap team is
calculated from Region 0 (Full sky with Galaxy cut) and then
corrected by using a maximum likelihood estimate \citep{wmap3ytem}. The power we have
calculated above, however, is directly from full-sky simulations
because one can then single out the effect of the MVO, whereas
extrapolation of the power from incomplete sky coverage may
introduce yet another systematic error, particularly for large
angular scales.

\section{Estimation of the quadrupole power}
Since we know the cosmic covariance is going to be subtracted by the
MVO, there is an unknown error in the quadrupole power even before
the cosmic variance becomes involved in the interpretation of the
data. Consequently, the quadrupole power we can  estimate is at best
not going to be a single value, but a sampling distribution which we
construct with help of Monte Carlo simulations, assuming the CMB is
a Gaussian random field. According to Table 1 (lower left triangle),
the full-sky \wmap foregrounds at K, Ka, Q and V channels (except W)
have correlation coefficients deviated from unity by less than $5
\times 10^{-3}$. We can employ the MVO on these 4 channels to retrieve the quadrupole power and then add on the missing power from Monte Carlo simulations.

\begin{figure}
\plotone{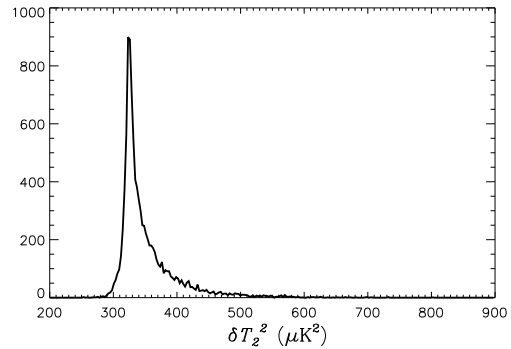}
\caption{Distribution of the CMB quadrupole power.
We employ the MVO on frequency band maps of the \wmap K, Ka, Q, V channels
(based on the excellent agreement in morphology between the foregrounds)
and add on the estimated estimated quadrupole the missing power from $10^4$ realizations of 4-channel MVO. The true CMB quadrupole falls in $[305.16,400.40] (\mu {\rm K}^2)$ (at the $1-\sigma$ level).}
\label{quadrupoledis}
\end{figure}

The weighting coefficients for the linear combination from the 4 frequency band maps 
are $(6.2419 \times 10^{-3}, -0.21597, -0.37970, 1.5894)$ and we denote the resultant map 
as ILC4. The correlation coefficients between the ILC4 and the derived foregrounds (frequency band maps subtracting ILC4) at K, Ka, Q and V are $ X=(8.3553, 8.3510, 8.3351, 8.2826) \times 10^{-8}$, respectively,
which confirms the presence of negligible correlations after the application of MVO.
The quadrupole power from the ILC4 map is $\delta T_2^2 = 326.30
\,\mu{\rm K}^2$. We then add the missing power for quadrupole from
$10^4$ realizations of exactly the same procedure as described
earlier in this Section, except this time using \wmap Q band foreground as common foreground for the 4 channels. In Fig.\ref{quadrupoledis} we plot the histogram of the resulting
estimates of true CMB quadrupole power, which falls in
$[305.16,400.40] (\mu {\rm K}^2)$ (at the $1-\sigma$ level) .

\section{Discussion and Conclusions}
We have shown that the MVO method for eliminating foreground
contamination from CMB maps is prone to a bias if there exist
accidental alignments between foreground features and structures in
the true CMB sky. This bias acts in such a way that a suppression of
the quadrupole is more likely than an enhancement. It also reduces
the overall variance of the recovered map compared to the true CMB
sky. This suggests that attempts to attribute the low quadrupole to
exotic physics may be premature.

The MVO was originally designed for the separation of Gaussian
signals, whose statistical properties are completely characterized
by their power spectra. For a linear combination of such signals,
the variance is clearly the natural choice for a parameter to be
minimized, and in such a case this method generally leads to a good
reconstruction of the power in each component. In this paper, we
have demonstrated even for the MVO to work on Gaussian signals it
requires a statistical ensemble: for a single realization the role
of cosmic covariance becomes crucial. The foregrounds relevant to
CMB analysis are also highly non--Gaussian, which makes this problem
even worse.

For composite foregrounds with varying spectra, the variance after
application of the MVO will induce even more foreground
contamination in the ILC map because the weighting coefficients will then be tuned
not only for the subtraction of cosmic covariance, but for varying
morphology among the different foregrounds. It is then possible that
the variance deficit generated by the subtraction of cosmic
covariance can be compensated by a foreground residual, which can
push the quadrupole power back to the value that is consistent with
the Concordance Cosmological model. Extra caution, therefore, will
have to be applied when employing the MVO to the interpretation of
results from the upcoming ESA Planck Experiment.

\smallskip

{\it Acknowledgments:} We acknowledge the use of \healpix
\footnote{{\tt http://healpix.jpl.nasa.gov/}}
package \citep{healpix} to produce $\alm$ from the \wmap data, and the use of \glesp\footnote{{\tt http://www.glesp.nbi.dk/}} package.


\vfill




\begin{thebibliography}{99}
\expandafter\ifx\csname natexlab\endcsname\relax\def\natexlab#1{#1}\fi
\newcommand{\combib}[3]{\bibitem[{#1}({#2})]{#3}} 

%
%
\newcommand{\autetal}[2]{{#1,\ #2., et al.,}}
\newcommand{\aut}[2]{{#1,\ #2.,}}
\newcommand{\saut}[2]{{#1,\ #2.,}}
\newcommand{\laut}[2]{{#1,\ #2.,}}

%
%
\newcommand{\refs}[6]{#5, #2, #3, #4} 
\newcommand{\unrefs}[6]{#5, #2 #3 #4 (#6)}  


%
%

\newcommand{\book}[6]{#5, #1, #2, #3}
%

\newcommand{\proceeding}[6]{#5, in #3, #4, #2} 

\def\apj{ApJ}
\def\apjl{ApJL}
\def\mn{MNRAS}
\def\nature{Nature}
\def\aa{A\&A}
\def\prl{Phys.\ Rev.\ Lett.}
\def\prd{Phys.\ Rev.\ D}
\def\pr{Phys.\ Rep.}
\def\ijmpd{Int. J. Mod. Phys. D}
\def\jcap{J. Cosmo. Astropar.}

\combib{Bennett~\etal}{2003b}{wmapresults}
\autetal{Bennett}{C. L}
\refs{First-Year Wilkinson Microwave Anisotropy Probe (WMAP)
  Observations: Preliminary Maps and Basic Results}
{\apjs}
{148}
{1}
{2003}
{astro-ph/0302207}

\combib{Bennett~\etal}{2003c}{wmapfg}
\autetal{Bennett}{C. L}
\refs{First-Year Wilkinson Microwave Anisotropy Probe (WMAP)
  Observations: Foreground Emission}
{\apjs}
{148}
{97}
{2003}
{astro-ph/0302208}

\combib{de Oliveira-Costa \& Tegmark}{2006}{dt}
\aut{de Oliveira-Costa}{A} \laut{Tegmark}{M}
\refs{CMB multipole measurements in the presence of foregrounds}
{\prd}
{74}
{023005}
{2006}
{astro-ph/0603369}

\combib{de Oliveira-Costa~\etal}{2004}{dtzh}
\aut{de Oliveira-Costa}{A} \aut{Tegmark}{M} \aut{Zaldarriaga}{M} \laut{Hamilton}{A}
\refs{The significance of the largest scale CMB fluctuations in WMAP}
{\prd}
{69}
{063516}
{2004}
{astro-ph/0307282}


\combib{Efstathiou}{2003b}{efsspacurv} 
\saut{Efstathiou}{G} 
\refs{Is
the Low CMB Quadrupole a Signature of Spatial Curvature?} {\mn}
{343} {L95} {2003} {astro-ph/0303127}



\combib{Efstathiou}{2003a}{efssig} \saut{Efstathiou}{G} \refs{The
Statistical Significance of the Low CMB Mulitipoles} {\mn} {346}
{L26} {2003} {astro-ph/0306431}

\combib{Efstathiou}{2004}{efsmla}
\saut{Efstathiou}{G}
\refs{A Maximum Likelihood Analysis of the Low CMB Multipoles from WMAP}
{\mn}
{348}
{885}
{2004}
{astro-ph/0310207}

\combib{Eriksen~\etal}{2004}{lilc}
\aut{Eriksen}{H K} \aut{Banday}{A J} \aut{G\'orski}{K M} \laut{Lilje}{P B}
\refs{On Foreground Removal from the Wilkinson Microwave Anisotropy
  Probe Data by an Internal Linear Combination Method: Limitations and
  Implications}
{\apj}
{612}
{633}
{2004}
{astro-ph/0403098}


\combib{G\'{o}rski, Hivon \& Wandelt}{1999}{healpix}
\aut{G\'{o}rski}{K. M} \aut{Hivon}{E} \laut{Wandelt}{B. D}
\proceeding{Analysis issues for large CMB data sets}
{PrintPartners Ipskamp, NL}
{A. J. Banday, R. S. Sheth and L. Da Costa}
{Proceedings of the MPA/ESO Cosmology Conference ``Evolution of
Large-Scale Structure''}
{1999}
{astro-ph/9812350}

\combib{Hinshaw~\etal}{2003b}{wmapcl}
\autetal{Hinshaw}{G}
\refs{First-Year Wilkinson Microwave Anisotropy Probe (WMAP)
  Observations: The Angular Power Spectrum}
{\apjs}
{148}
{135}
{2003}
{astro-ph/0302217}



\combib{Hinshaw~\etal}{2007}{wmap3ytem}
\autetal{Hinshaw}{G}
\refs{Three-Year Wilkinson Microwave Anisotropy Probe (WMAP) Observations: Temperature Analysis}
{\apjs}
{170}
{288}
{2007}
{astro-ph/0603451}

\combib{Luminet et al.}{2003}{dodecahedral}
\aut{Luminet}{J.-P} \aut{Weeks}{J} \aut{Riazuelo}{A} \aut{Lehoucq}{R} \laut{Uzan}{J.-P}
\refs{Dodecahedral space topology as an explanation for weak wide-angle temperature correlations in the cosmic microwave background}
{\nat}
{425}
{593}
{2003}
{astro-ph/0310253}

\combib{Monteserin et al.}{2007}{lowvar}
\autetal{Monteserin}{C} 
\unrefs{A low CMB variance in the WMAP data}
{\mn}
{submitted}
{}
{2007}
{arXiv:0706.4289}

\combib{Park, Park \& Gott}{2007}{park3y}
\saut{Park}{C.-G}  \aut{Park}{Changbom} \laut{Gott}{J R III}
\refs{Cleaned Three-Year WMAP CMB Map: Magnitude of the Quadrupole and Alignment of Large Scale Modes }
{\apj}
{660}
{959}
{2007}
{astro-ph/0608129}

\combib{Saha et al.}{2007}{saha}
\aut{Saha}{R} \aut{Prunet}{S} \aut{Jain}{P} \laut{Souradeep}{T} 
\unrefs{CMB anisotropy power spectrum using linear combinations of WMAP maps}
{\prd}
{submitted}
{}
{2007}
{arXiv:0706.3567}

\combib{Tegmark}{1998}{realworld}
\saut{Tegmark}{M}
\refs{Removing Real-World Foregrounds from Cosmic Microwave Background Maps}
{\apj}
{502}
{1}
{1998}
{astro-ph/9712038}


\combib{Tegmark, de Oliveira-Costa \& Hamilton}{2003}{toh}
\aut{Tegmark}{M} \aut{de Oliveira-Costa}{A} \laut{Hamilton}{A}
\refs{A high resolution foreground cleaned CMB map from WMAP}
{\prd}
{68}
{123523}
{2004}
{astro-ph/03022496}


\combib{Tegmark \& Efstathiou}{1996}{te96}
\aut{Tegmark}{M} \laut{Efstathiou}{G}
\refs{}
{\mn}
{281}
{1297}
{1996}
{}


\combib{White, Krauss and Silk}{1993}{cosmicvariance}
\aut{White}{M} \aut{Krauss}{L M} \laut{Silk}{J}
\refs{Cosmic Variance in Cosmic Microwave Background Anisotropies: From 1 degrees to COBE}
{\apj}
{418}
{535}
{1993}
{astro-ph/9303009}





\end{thebibliography}
\end{document}